\documentclass[sigconf]{acmart}
\usepackage{tikz}
\usepackage{xcolor}
\usetikzlibrary{positioning, shapes, arrows.meta}

\AtBeginDocument{%
  \providecommand\BibTeX{{%
    \normalfont B\kern-0.5em{\scshape i\kern-0.25em b}\kern-0.8em\TeX}}}

\copyrightyear{2024}
\acmYear{2024}
\setcopyright{acmlicensed}\acmConference[DIS '24]{Designing Interactive Systems Conference}{July 1--5, 2024}{IT University of Copenhagen, Denmark} \acmBooktitle{Designing Interactive Systems Conference (DIS '24), July 1--5, 2024, IT University of Copenhagen, Denmark} \acmDOI{10.1145/3643834.3661507} \acmISBN{979-8-4007-0583-0/24/07}

\include{Chapters/_macros}
\usepackage{spverbatim}

\begin{document}

\title[The CoExplorer Generative AI Adaptive Meeting UI]{The CoExplorer Technology Probe: A Generative AI-Powered Adaptive Interface to Support Intentionality in Planning and Running Video Meetings}

\author{Gun Woo (Warren) Park}
\email{warren@dgp.toronto.edu}
\affiliation{
\institution{Microsoft Research}
  \city{Cambridge}
  \country{United Kingdom}
}
\affiliation{%
  \institution{University of Toronto}
  \city{Toronto}
  \country{Canada}
}

\author{Payod Panda}
\authornote{Corresponding author}
\email{payod.panda@microsoft.com}
\affiliation{%
  \institution{Microsoft Research}
  \city{Cambridge}
  \country{United Kingdom}
}

\author{Lev Tankelevitch}
\email{lev.tankelevitch@microsoft.com}
\affiliation{%
  \institution{Microsoft Research}
  \city{Cambridge}
  \country{United Kingdom}
}

\author{Sean Rintel}
\email{serintel@microsoft.com}
\affiliation{%
  \institution{Microsoft Research}
  \city{Cambridge}
  \country{United Kingdom}
}

\renewcommand{\shortauthors}{Park and Panda, et al.}

\begin{abstract}

Effective meetings are effortful, but traditional videoconferencing systems offer little support for reducing this effort across the meeting lifecycle. Generative AI (GenAI) has the potential to radically redefine meetings by augmenting intentional meeting behaviors. %
CoExplorer, our novel adaptive meeting prototype, preemptively generates likely phases that meetings would undergo, tools that allow capturing attendees’ thoughts before the meeting, and for each phase, window layouts, and appropriate applications and files. Using CoExplorer as a technology probe in a guided walkthrough, we studied its potential in a sample of participants from a global technology company. Our findings suggest that GenAI has the potential to help meetings stay on track and reduce workload, although concerns were raised about users’ agency, trust, and possible disruption to traditional meeting norms. We discuss these concerns and their design implications for the development of GenAI meeting technology.

\end{abstract}



\begin{teaserfigure}
  \includegraphics[width=\textwidth]{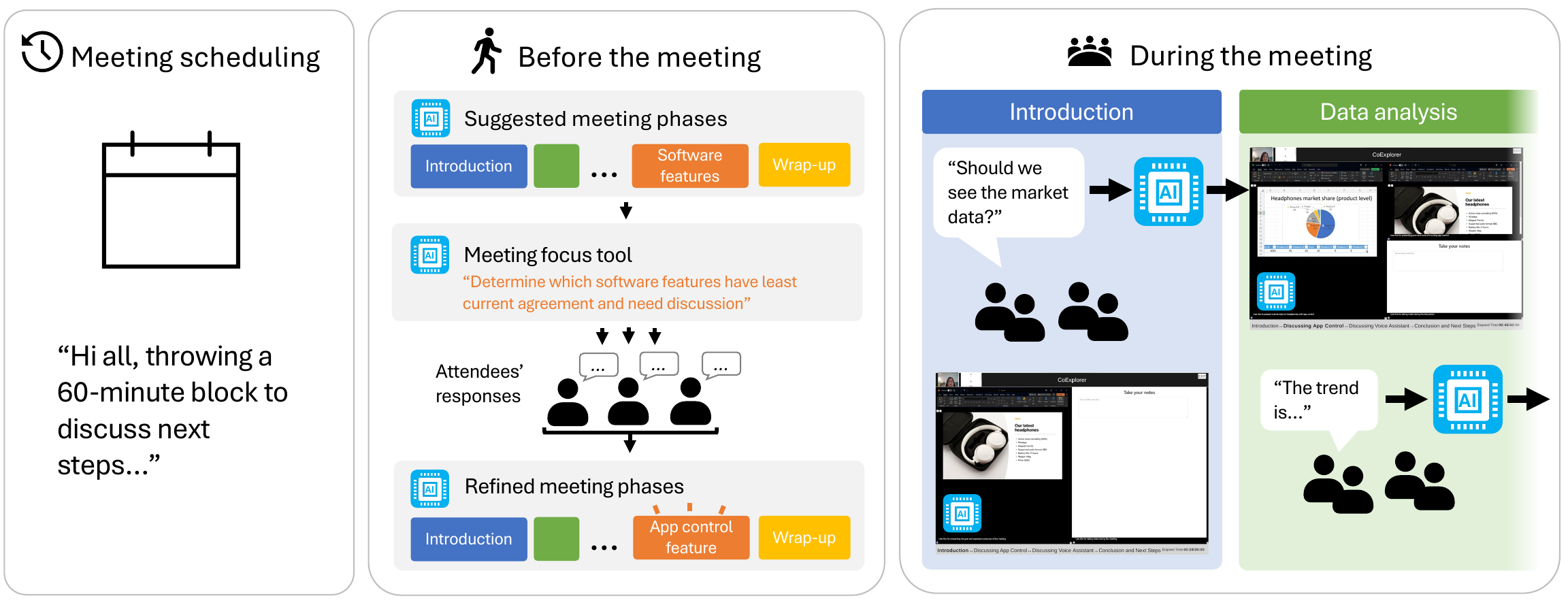}
  \caption{CoExplorer uses Generative AI to reduce the effort of meeting intentionality. From a natural language calendar invitation, GenAI is used to generate a meeting goal, a suggested list of meeting phases and their associated activities, and a meeting focus tool to help find specific needs for discussion. From attendee responses to that tool, GenAI refines the list of meeting phases. When the meeting is underway, GenAI is used to monitor the talk and either suggest or follow users' transitions between phases to fit the trajectory of the meeting. For each phase, GenAI generates apps and optimizes their position and size to fit the meeting phase.}
  \Description{Shows the process in CoExplorer. The first step is meeting scheduling, in which the meeting invitation text is defined. The second is before the meeting, during which CoExplorer (1) suggests meeting phases, (2) generates a meeting focus tool to gather attendees' responses, and (3) refines meeting phases. The last step is during the meeting, in which CoExplorer changes the window layout using the layout generated by AI based on the detected phase, based on the speech of the attendees.
  }
  \label{fig:teaser}
\end{teaserfigure}

\received{8 February 2024}
\received[accepted]{18 April 2024}

\maketitle

\section{Introduction}
\label{sec:introduction}

Video meetings have enabled a new era of distributed work, but this has not gone hand-in-hand with improved meeting effectiveness~\cite{microsoft_work_2023,lucid_howmanymeetings_2022,doodle__meetingsgeneral_2019}, and may have introduced additional  fatigue~\cite{bailenson_nonverbal_vcfatigue_2021}.
Evolving from the telephony technology paradigm~\cite{kraut_prospectsforvideotelepphony_1997,crowcroft_videoconfontheinternet_1997}, videoconferencing systems have largely focused on creating a connection between endpoints and representing video and audio on a canvas.
While research has explored broadening the technological support for person, reference, and task spaces~\cite{buxton_mediaspace_2009}, scheduling~\cite{chun_optimizing_meetingscheduling_2003}, agendas~\cite{garcia_voting_2005}, moderation~\cite{aseniero2020meetcues}, and even decision-making ~\cite{hiltz_gss_2015}, improving holistic meeting effectiveness by reducing the effort of planning and running meetings remains an unmet design challenge.
For example, calendar interfaces make it easy to schedule a meeting, but very few require the user to specify a meeting \textit{goal}, and videoconferencing interfaces tend to simply provide a place for repeating the text of invitations, without making use of that information as relevant to the interface design or experience~\cite{scott_2024_mentalmodelsmeetinggoals}.

``Generative AI'' (GenAI) is a generic term encompassing all end-user tools whose technical implementations include a generative model based on deep learning~\cite{sarkar-2023-genai-definition}. GenAI has the potential to solve these problems and open up a new intentionality-driven design paradigm for video meetings.  
Commercial videoconferencing systems such as Zoom Companion AI~\cite{taylor_zoom_aimeeting_2023} and Microsoft Teams Meeting Copilot~\cite{microsoft_copilot_aimeeting_2023} are already building GenAI features into their meeting systems.
However, these just scratch the surface of how GenAI could help teams have more intentional meetings and overcome long-standing ineffectiveness.

This paper reports on a study of the design opportunities and challenges in this space through CoExplorer, a GenAI-powered technology probe~(see \autoref{sec:coexplorer}). 
CoExplorer represents some of the core capabilities of a holistic goal-driven adaptive system for planning and running meetings. 
CoExplorer: (1) uses a meeting invitation to generate tentative meeting goal and phases, (2) uses the meeting goal to generate a meeting focus tool, (3) suggests an appropriate set of applications and their layout for phases (detected in real-time) during a meeting, and (4) analyzes meeting conversation to fine-tune the phase structure automatically.

We conducted a user study involving 26 participants from a global technology company. 
In the hour-long study, participants were given a guided tour of CoExplorer, frequently engaging with the researcher to discuss the pros and cons of design concepts across the meeting lifecycle. %
We inquired about their views on various system elements, including their benefits, drawbacks and potential issues. 
Our thematic analysis of the interviews found that many participants valued the system's ability to facilitate attendee alignment on meeting purpose, and to automatically choose relevant applications or files based on the current meeting phase. 
However, they also emphasized the importance of verifying the system's choices and appreciated prompts for confirmation at appropriate intervals. 
These findings highlight an underlying conflict between ease-of-use and higher agency, and we discuss strategies to balance these tensions in our design implications. 
Our contributions are:
\begin{itemize}
    \item CoExplorer, a GenAI-powered meeting system that optimizes task space in response to meeting goal-relevant input and immediate activity context,
    \item Findings from a user study that sheds light on the perceived merits, utility, and concerns of our novel system
    \item Design insights for future creators interested in developing adaptive windowing systems
    \item A set of system and user prompts that could facilitate the development of a similar system
\end{itemize}

\section{Background and Related Work}
\label{sec:relatedwork}

\subsection{Meetings and effectiveness}

Planning and preparing for an effective meeting takes effort. Much effort goes into appropriate scheduling given different preferences and relevance of attendees~\cite{sen_formal__meetingscheduling_1998,mok_timezone_timepressure_2023,gibbs_work_timepressure_2021,chun_optimizing_meetingscheduling_2003}. 
However, the larger problem is whether the meeting should occur at all and whether it has clear goals. 
Clear goals are imperative for effective meetings~\cite{bang_effectiveness_2010,geimer_meetings_meetingsgeneral_2015}, but setting goals takes effort and time of an organizer and potentially attendees~\cite{garcia_agenda_voting_2004}, which many aspects of organizational culture may work against. 
There is often a lack of understanding of the kind of tasks best suited to meetings~\cite{martin_state_timepressure_2020,qatalog_killingtime_2023,bergmann_meeting__vcfatigue_2022,microsoft_momentsthatmatter_2023}. 
Even when goals are set, meeting preparation is often sacrificed due to time pressures~\cite{geyer_directors__meetingprep_2020,elsayedelkhouly_thirdoftimewasted_meetingprep_1997}.

Running and participating in an effective meeting also takes effort. 
Agendas are a valuable tool in structuring meetings~\cite{cutler_meeting_2021, garcia_agenda_voting_2004}, as long as they do not become rote or a crutch~\cite{rogelberg_surprising_meetingsgeneral_2019}. 
One important aspect is managing the agenda while also allowing for flexibility necessary to incorporate new ideas, disagreement etc. ~\cite{kocsis_effectivemeetings_2015,rogelberg_surprising_meetingsgeneral_2019,usmani_impact_2012, koshy_how_2017}. 
Agendas are often a source of tension in terms of how many topics will be included and who wants those topics raised, which leads to the question of which topics must be discussed versus which could be resolved asynchronously before the meeting to condense the agenda and leave time for discussion of differences or productive conflict~\cite{garcia_voting_2005,qatalog_killingtime_2023,miranda_meetingconflict_gss_1993}.
While the agenda provides a topical structure to manage, the other effort of running a meeting is the work of managing the meeting's \textit{phases}~\cite{deppermann_agenda_2010}. 
Phases are organized locally in the moment, and may be synonymous with agenda items, be more granular than an agenda item, or encompass several agenda items. 
Phases are activity-oriented, often involving the use of multiple resources, and different phases may require different resources. 
Here the effort is in ensuring that the relevant participants are focused on the relevant materials at the right time.

Post-meeting effort starts during the meeting in capturing notes and toward the end capturing action items or next steps~\cite{rogelberg_surprising_meetingsgeneral_2019,moran_savlagingmeetingrecords_postmeeting_1997,kalnikaite_markupasyoutalk_postmeeting_2012,morgan_actionitems_postmeeting_2009}. 
Significant research has gone into reducing this effort through automatic generation of meeting summaries, highlights, and action items~\cite{asthana_summaries_postmeeting_2023}. 
If a meeting has unclear goals and/or not been well-run, post-meeting effort may be higher or another meeting may be required, creating a vicious circle of effort~\cite{rogelberg_surprising_meetingsgeneral_2019}.

In this paper we focus on planning and running meetings as the two most crucial points of intervention. 
The effort of planning and running meetings apply in both unmediated and mediated situations, but video-mediated meetings have added layers that either do not support the effort or even exacerbate it.
There is a need for adaptive video meeting systems to reduce the effort of planning by helping turn goals into meaningful action phases for the meeting, and then starting and flexibly changing phases to reduce the effort of running and participating in the meeting.

\subsection{Using AI to Improve the Effectiveness and Adaptability of Video Meetings}

AI interventions have been proposed as a solution to many kinds of meeting problems, such as to reduce the effort of \textit{pre-meeting} work such as automating meeting scheduling~\cite{chun_optimizing_meetingscheduling_2003} and agenda item voting~\cite{garcia_voting_2005, garcia_agenda_voting_2004,garcia_agenda_voting_2003}. 
Research has also focused on improving decision-making \textit{during the meeting} itself. 
The field of Group Support Systems (GSS) was very large in the 1980s through early 2000s~\cite{hiltz_gss_2015}. 
De Vreede et al.~\cite{devreede_gssfailed_2003}, report that the most common reason for failure in meetings relying on Group Support Systems was "Goal[s] poorly defined by the process owner"---yet ironically, direct technological support for expressing goals was often lacking in such systems. 
Since then, attention has shifted to how AI agents can facilitate various aspects of meetings, often focused on automatically detecting people's roles \cite{banerjee2004using,vinciarelli2011understanding}, useful actions~\cite{mcgregor2017more}, and social dynamics \cite{bhattacharya2018multimodal} in meetings, and on issues of inclusion and participation (e.g. ~\cite{guo_agents_nonnative_2019,kohl_vizcontribs_2023,fu2022improving, muller2018detecting}) than directing the meeting's purpose. To the latter point, \citet{kim2016improving} showed that a model that automatically assesses consistency of understanding across meeting participants and suggests topics for review has potential to improve team understanding. 
\citet{aseniero2020meetcues} showed that a system enabling participants to provide real-time meeting feedback via a back-channel can improve their engagement and awareness. 
Other work showed that automatic detection of different types of conversations \cite{zhou2021role}, emotions \cite{zhou2022predicting}, bodily cues \cite{choi2021kairos}, and environmental factors \cite{constantinides2020comfeel} in meetings can predict aspects of meeting effectiveness.
For reducing \textit{post-meeting} effort, significant research has gone into automatic generation of meeting summaries, highlights, and action items,  ~\cite{asthana_summaries_postmeeting_2023,moran_savlagingmeetingrecords_postmeeting_1997,kalnikaite_markupasyoutalk_postmeeting_2012,morgan_actionitems_postmeeting_2009}, as well as metrics \cite{cutler_meeting_2021,hosseinkashi2024meeting} and post-meeting dashboards that attempt to provide feedback on meeting effectiveness~\cite{samrose_collabcoach_2018,samrose2021meetingcoach}.
GenAI has sparked a surge in commercial meeting assistance tools.  
Zoom Meeting AI Companion~\footnote{https://news.zoom.us/zoom-ai-companion/} and Microsoft Teams Meeting Copilot~\footnote{https://support.microsoft.com/en-us/copilot-teams} build AI tools directly into videoconferencing, while Read~\footnote{https://www.read.ai/} and Fellow~\footnote{https://fellow.app/} are examples of AI services that can integrate with a range of videoconferencing applications.

Both the prior research and commercial tools show promise, but they have shortcomings if we consider the overall problems of ineffective meetings and meeting fatigue. 
First, they concentrate on features to improve specific pain points, usually because there is a clear technical path to a solution (e.g. automated agenda creation). 
However, in some cases these features \textit{add} to effort, such as providing flexibility that needs to be manually managed (e.g. using a meeting focus tool~\cite{garcia_voting_2005} or choosing window layouts~\cite{houben_co-activity_2012}). 
More importantly, they miss the bigger picture of improving the \textit{intentionality} of meetings across the meeting lifecycle---everyone in a meeting knowing why they are there, what it is for, and what they need to discuss. In sum, the research gap we explore is how to design a system that reduces the overall effort of meeting intentionality through a goal-driven approach to \textit{focusing} the communication, refinement, and moderation of purpose-driven effort, and \textit{eliminating} technical overhead in organizing, moderating, sharing documents, and arranging the user interface.

\subsection{Adaptive User Interfaces}

Adaptive user interfaces adapt to varying user needs either at the time of setup and/or dynamically during operation. 
By adapting to in-the-moment user needs, adaptive interfaces reduce cognitive load by improving navigability of information \cite{brusilovsky_methods_1996, findlater_ephemeral_2009, todi_adapting_2021, matejka_patina_2013, al-omar_user_2009, shankar_user-context_2007, gajos_predictability_2008}. 
One of the simplest adaptive interfaces is context-dependent adaptive windowing systems. 
These enable easy retrieval of groups of tools needed in the moment, either as groups of windows per task \cite{tashman_windowscape_2012, tashman_windowscape_2006, smith_groupbar_2003} or time \cite{hu_scrapbook_2020}. 
Co-Activity Manager~\cite{houben_co-activity_2012, houben_activity-centric_2013} combines traditional window configuration task-switching with the ability to share a particular configuration with collaborators. 
In such systems, users manually choose window configurations, although sometimes the placement of components may be optimized automatically~\cite{park_adam_2018}. %
AI has obvious application for optimizing adaptive windowing of relevant resources to fit with context and/or device affordances~\cite{oulasvirta_combinatorial_2020, stefanidi_context-aware-ui-example_2022, stige_ai-user-experience-slr_2023}.

In videoconferencing contexts, the layout of interfaces has received attention in specific areas. 
The issue of whether video adds value to audio at all is a  classic problem, and a recent study has found that video of content is important for all business needs video of content is needed but not necessarily video of people~\cite{standaert_how_layout_2021}. 
When people are considered to be important, much attention has been paid to exotic solutions that preserve eye gaze, usually by warping the image of the eyes~\cite{jerald_gaze_warping_2002} but sometimes by changing the interface itself ~\cite{vertegaal_gaze2_layout_2003}. 
Beyond gaze, preserving spatiality has also been explored~\cite{tang_perspectives_layout_2023,hauber_spatiality_layout_2006}. 
End-users also configure their window layouts  to suit their working preferences as best they can given the limited flexibility of videoconferencing applications~\cite{karolina_configurezoom_layout_2022}, and some classes of end-users (e.g. no- and low- vision~\cite{ang_signers_layout_2022, tang_disabilities_layout_2021}) find that the videoconferencing interface was never designed to suit their needs.

However, \textit{adaptive interfaces} have had limited exploration. 
One long-standing limitation in videoconferencing has been how to enable remote users to see enough of one another or an environment to make sense of a task in context~\cite{junuzovic_idealayouts_adaptive_vc_2011}. 
Research has investigated various approaches to fluid group formation~\cite{hu_fluidmeet_adaptive_vc_2022}, moving video around to improve reference-space capabilities~\cite{gronbeck_mirrorblender_adaptive_vc_2021} and automated view direction or focus logic~\cite{weiss_orchestrator_adaptive_vc_2014,yao_focalspace_adaptive_vc_2013}.

We have found very little on adaptive approaches applied to the agenda or phases of meetings, nor to the display and arrangement of task-oriented materials in video meetings. 
One example is a 1995 attempt at adapting interface layouts to work contexts. 
The "dynamically adaptive multi-disciplinary workstation" (DAMDW)~\cite{chimiak_telemedicine_adaptive_1995} changed its display of tools at the outset of a telemedicine consultation to suit a physician's specialization. 
However, it did not change thereafter and could not adapt to context changes in the moment. 
In 2003, Antunes and Costa~\cite{antunes_meetingwaredesign_meetinglifecycle_2003} investigated whether genre analysis could be used not just to categorize meetings and activities, but encoded into an interface to reflect the needs of a meeting. 
They found that the genre concept was useful to clarify the organizational context of meetings and that the systems they tested supported recurrent work practices without undue constraint. 
However, writing in 2003, they also found that doing the work manually was not a scalable approach: "we currently cannot envisage a way to generalize so many different meetings occurring in organizations." 
More recently, VisPoll~\cite{chung_scalablevisual_adaptive_vc_2021} enables an education video streamer to set up areas of a visual stream that users may interact with, which can be used to enable a large audience to provide visual input to questions asked about the stream, and the streamer, in turn, to manage their cognitive load by seeing the aggregate of those audience inputs in a manageable visual fashion. 
However, this system does not use AI and requires effort of the streamer to decide on the visual polling to be used.

GenAI systems using Large Language Models (LLMs) are well suited to meeting interventions because they are very good at performing using natural language inputs. 
Text is ideal for pre-meeting requirements, and during the meeting a Voice User Interface (VUI) is well-suited to driving textual representation into language models\cite{dasgupta_voice_2018}. 
Proactive VUIs~\cite{zargham_understanding_2022, wei_understanding_2021, schmidt_user_2020} can support time-critical events, which is essential for recognizing how well meeting talk matches meeting needs.
Researchers have utilized LLMs as a tool for prototyping decision-making algorithms in interactive system prototypes~\cite{pu_semanticon_2022, li_stargazer_2023}, but less so in adaptive meeting systems. One example is %
CrossTalk~\cite{xia_crosstalk_2023}, which harnesses an intelligent substrate to prompt context-appropriate controls such as screen sharing or to present pertinent information. %
Another inspiratino is Co-activity Manager~\cite{houben_co-activity_2012}, which combines a context-dependent windowing system, a shared task space, and critically, an intelligent adaptive interface. 

We speculate that the sparsity of adaptive interface exploration for meetings has to do with the difficulty of balancing flexibility with the manual work required to actually make meaningful changes on the fly, especially if that distracts users from carrying out the actual tasks of the meeting. 
GenAI holds the promise of reducing this manual load~\cite{park2024coexplorer}. 
That being said, using GenAI to enable such capabilities brings its own challenges, among which is that predictions are non-deterministic. 
This creates risks of users being unable to achieve tasks as expected or in repeatable ways~\cite{dove_ux_2017}.

\subsection{Research questions}

In sum, video meetings are essential for collaborative work and decision-making, but have been largely stuck in a one-size-fits-all container optimized for connection without much intentionality. 
The status quo leaves goal setting, agenda creation, and scene setting to people who are often too busy to prepare before the meeting. 
It also leaves meeting facilitation and resource management to people who are already under cognitive load making decisions about the content and potentially experiencing videoconferencing fatigue based on the interface. 
To that end, this paper uses a technology probe to explore how GenAI may offer more holistic support for video meetings. 
We explore three research questions about the opportunities and challenges that users face with a video meeting system that uses GenAI:

\begin{enumerate}
    \item[RQ1.] What are users' perceptions of GenAI driving a meeting's purpose for a collaborative meeting of teammates?
    \item[RQ2.] What are users' perceptions of GenAI surfacing implicit needs and resources in meetings, and the potential effects on participation and effectiveness?
    \item[RQ3.] What are users' perceptions of human-on-the-loop interactivity in GenAI video meeting systems?

\end{enumerate}

These questions drive the design concepts that we implement in our technology probe, as laid out in \autoref{sec:designconcepts}, our findings in \autoref{sec:findings}, and our discussion in \autoref{sec:discussion}.

\section{Designing CoExplorer}
\label{sec:coexplorer}

We began by choosing a meeting type that would benefit from GenAI-augmented help with planning and running a meeting, getting motivation from existing work e.g.,~\cite{park2024coexplorer}. 
We landed on meetings of cross-functional product teams in the technology industry, which often face struggles of coordination~\cite{jassawalla_crossfunctionalteams_2000}, especially at the start of projects when they need to form plans about which they have different opinions and stakes, and for which the resources are scattered across different storage locations and apps. We then developed a fictional meeting scenario emphasizing the need for effective decision-making. In this scenario, the team's current product, the Strata Headphones 2, is lagging behind competitors in market share, and so the team (hardware and software developers, designers, and researchers) needs to decide on the feature set that will increase market share of a new version. The team has files such as the list of design, hardware, and software features that can be included, the price for implementing them, competitor information, and the current product's specification sheet.

\begin{figure*}[!htb]
    \centering
    \includegraphics[width=\textwidth]{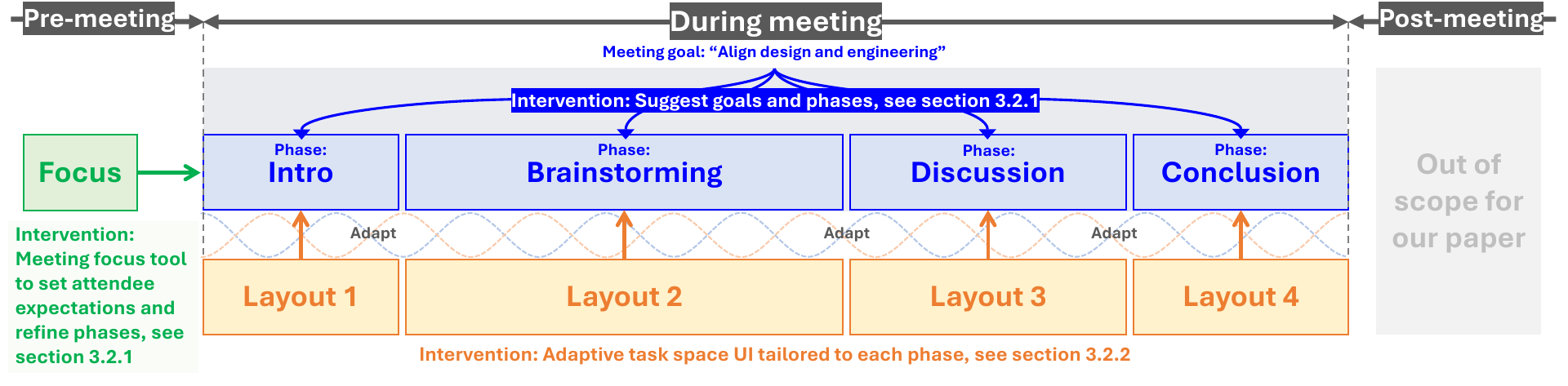}
    \caption{This framework shows the pre-, during, and post- meeting stages and identifies pain points (initiating focused discussion; resource management for each phase), and suggests points of intervention (meeting focus tool; adaptive phase definitions; layouts tailored to each phase).}
    \Description{Shows the breakdown of the meeting. There are pre-meeting, during-meeting, and post-meeting stages. The post-meeting stage is outside of our scope.}
    \label{fig:framework}
\end{figure*}

\subsection{Design Strategies for GenAI interventions}
\label{sec:designconcepts}

After defining the scenario and design concepts, we considered where CoExplorer would provide the most meaningful interventions in the meeting lifecycle. One early decision was to leave the post-meeting stage out of this probe, partially because there has been significant work on meeting summarization and dashboards already (as reviewed e.g., ~\cite{asthana_summaries_postmeeting_2023,moran_savlagingmeetingrecords_postmeeting_1997,kalnikaite_markupasyoutalk_postmeeting_2012,morgan_actionitems_postmeeting_2009, samrose_collabcoach_2018,samrose2021meetingcoach}) and partially because, as above, the level of post-meeting effort is dependent on the planning and running of the meeting. As such, we concentrated on the pre- and during meeting phases. \autoref{fig:framework} below identifies the primary pain points in these phases (initiating focused discussion; resource management for each phase), and forms of intervention (meeting focus tool; adaptive phase definitions; layouts tailored to each phase). \autoref{fig:framework} can be read in conjunction with \autoref{fig:teaser}, which shows a timeline of how meeting attendees would experience CoExplorer. We used this framework to craft a narrative of how CoExplorer would provide interventions in the scenario.

\textit{\textbf{Design Strategy 1: Clarify underlying needs and available resources.}}
The clear definition of objectives, agendas, preparatory materials, and task-related resources is an essential aspect of efficient meeting preparation and facilitation, notably in virtual settings \cite{cohen_meeting_2011, kreamer_optimizing_2021}.
Yet, due to time constraints, both meeting planners and participants may cut corners in laying out pre-meeting preparations \cite{scott2024goals}.
During the meeting, while \textit{explicit} items on the agenda direct the discourse, the meeting actually often transitions through \textit{implicit} phases.
These phases may align with agenda points but can also encompass several points or be subdivided into finer details.
Moreover, essential files and applications are often linked to each action-oriented phase. The implicit nature of these phases makes accessing appropriate resources at the opportune moment burdensome.
This design strategy explores the challenges and potential benefits of GenAI in identifying implicit phases as they arise during the meeting, and using them to drive a specific arrangement of resources and the meeting’s intended progression.

\textit{\textbf{Design Strategy 2: Incorporate collective feedback to shape meeting objectives.}}
Agendas are typically set by meeting organizers, but should also reflect the priorities of other attendees \cite{svennevig_agenda_2012}, yet current videoconferencing systems do not easily allow for collaborative crafting of meeting objectives. %
Participants’ contributions can be integrated through mechanisms like voting systems to create or fine-tune the agenda~\cite{garcia_voting_2005}.
This design strategy explores the potential of GenAI to produce a tool to assimilate varying viewpoints, focusing the conversation on reconciling differences.

\textit{\textbf{Design Strategy 3: Manage the system through a \textit{Human-on-the-Loop} (HOTL) methodology.}}
HOTL characterizes human-machine interactions in which the automated system mainly allows humans to \textit{abort} the machine's decisions, once initiated by the system~\cite{nahavandi_trusted_2017}. This is in contrast to \textit{Human-in-the-Loop} (HITL), where the system requires user approval \textit{prior to} executing an action.
This grants automated systems more independence and limits the number of prompts to users, which is compatible with a meeting context where participants often have lower capacity to process information outside the interpersonal interactions of meetings \cite{ofek_reducing_2013}.
Introducing HOTL in meeting systems has been impeded by inadequate predictive performance in natural language models, but with the launch of GPT3.5, there has been noticeable improvement in accuracy across a spectrum of fields \cite{johnson_assessing_2023, chen_how_2023}.
Nonetheless, to foster and sustain the trust of users, HOTL systems must clear a higher threshold than the often-preferred human-\textit{in}-the-loop methodology in Human-Computer Interaction \cite{xia_crosstalk_2023, pu_semanticon_2022, dong_webrobot_2022,nahavandi_trusted_2017}.
This design strategy explores the difficulties and possibilities of building trust with users through HOTL, and how this might influence user perceptions of both efficiency and adaptability.

\textbf{Overall implementation:}
We implemented CoExplorer using Unity and the GPT-3.5 and GPT-4 GenAI models. 
We utilized LiveKit for real-time video communication to provide cross-platform compatibility.
With CoExplorer, we wanted to explore the extent to which GenAI could generate appropriate meeting interfaces--
accordingly, we implement our design strategies primarily through designing GenAI assistants that generated the visual interfaces satisfying our requirements.
For implementation details, we include all our system prompts in~\autoref{sec:appendix-B}, and specific details for the various strategies in the section below.

\begin{figure*}[th!]
    \centering
    \includegraphics[width=\textwidth]{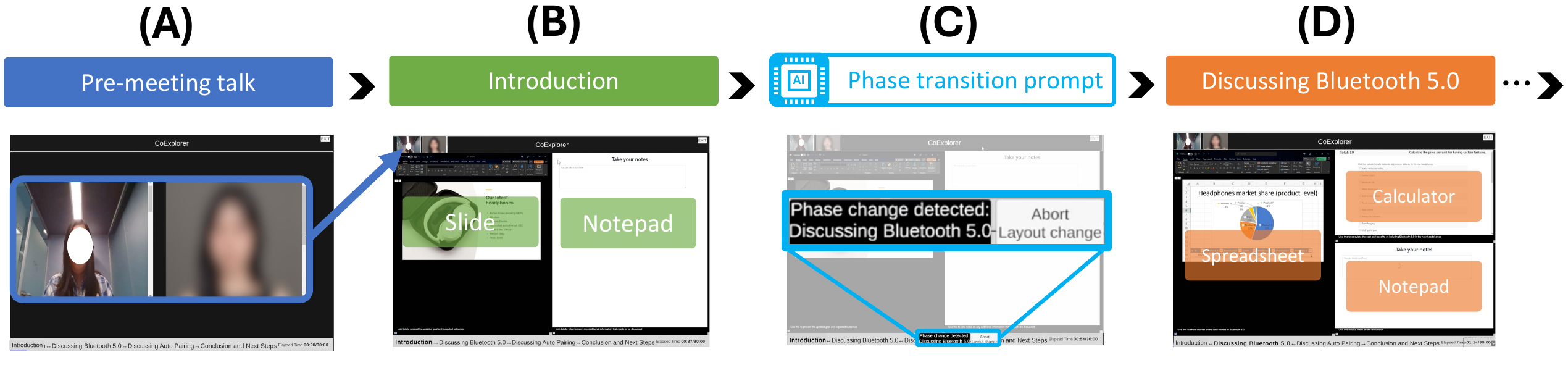}
    \caption{Upon recognition of a phase transition, CoExplorer notifies users and modifies the display to accommodate the new phase.}
    \Description{Shows how window layouts get changed depending on the phase. (A) when attendees are in a pre-meeting talk, attendees see a video of each other. (B) When they move into the introduction phase, slide and notepad windows come up, while minimising the videos. (C) When the system detects the meeting phase transition through AI, a phase transition prompt appears which allows users to abort the transition. (D) Phase change happens if the phase change is not aborted. Shows spreadsheet, calculator, and notepad windows.}
    \label{fig:windowingSystem}
\end{figure*}

\subsection{How CoExplorer Facilitates Meetings}

\begin{figure*}[ht!]
    \centering
    \includegraphics[width=0.8\textwidth]{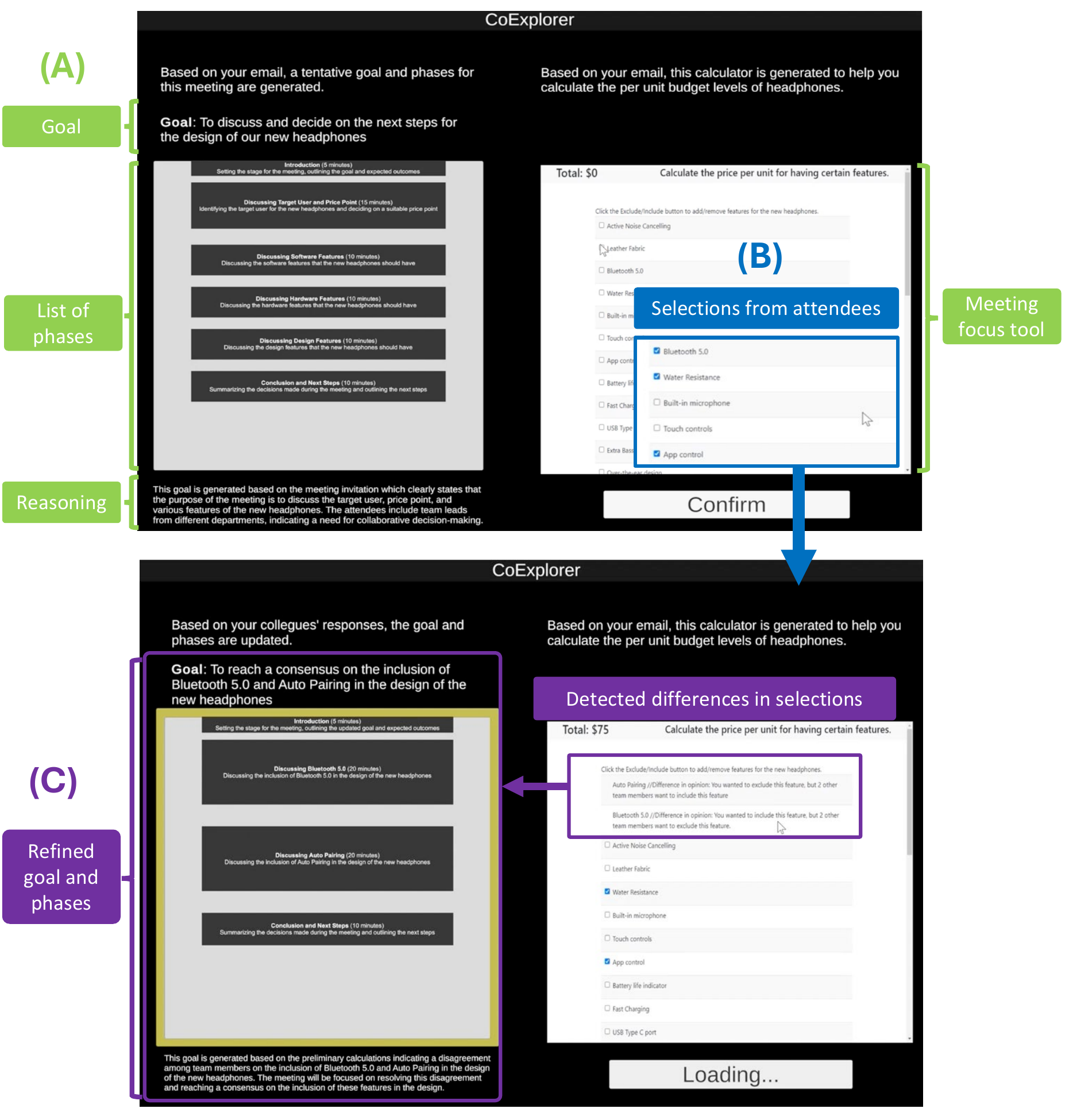}
    \caption{(A) Sequence of phases and tool for determining meeting focus. (B) Participants in the meeting employ the focus tool to select preferred features. (C) CoExplorer uses the aggregated preferences to adjust the meeting's objective and flow.}
    \Description{Shows screens in CoExplorer. The first screen shows the (A) goal, list of phases, and reasoning on the left-hand side, and the generated meeting focus tool on the right-hand side. (B) When the attendees make a selection on the meeting focus tool, (C) goals and phases get refined, based on the detected differences in selection}
    \label{fig:init}
\end{figure*}

\subsubsection{Formulating the Initial Phases of the Meeting}
The meeting organizer distributes a meeting invitation, and as the meeting time approaches, for each attendee CoExplorer outlines the meeting's goals (\autoref{fig:init}A, top left) and describes the reasoning for those goals (\autoref{fig:init}A), bottom left).
It also suggests phases anticipated for the meeting (\autoref{fig:init}A), left, shown as segmented bars (see~\autoref{sec:appendix-A},~\autoref{fig:phasesSpecific}, left).
Details for each phase include the name of the phase, the expected duration, and the pertinent activities for that phase.
Below we expand on how the system implements the design strategies above. The strategies are not implemented in the order presented, and one strategy might be implemented multiple times.

\textbf{Implementing Design Strategy 1:} Inspired by the successful use of chaining in LLMs~\cite{wu_promptchaining_2022}, we posited that given a brief meeting invitation text, GenAI could identify both the purpose of the meeting, and a tailored list of applications for each phase.
We emphasize our focus is on exploring the potential applications enabled by GenAI's capabilities, and thus formal evaluation of this capability is out of scope for this paper.
GenAI is tasked with elaborating on the provided invitation text and compiling a list of the phases of the meeting, complete with titles and descriptions.
We then guide GenAI to structure these details into a coherent list that includes explanations for the choices it has made, crafting a narrative that can both inform the CoExplorer system and persuade users of the validity of the decisions.
To achieve this, we initialise GenAI requests with system prompts, bolstered by examples, to ensure the output is clear and actionable (see Appendix \ref{sec:appendix-B-phase} for the system prompts for phase generation.
This preliminary information generated by GenAI is fixed at the beginning and does not undergo refinement during the meeting, intentionally designed to preclude any confusion that might arise from changes during the meeting.

\textbf{Implementing Design Strategy 2:} CoExplorer also primes participants to consider their own requirements and those of the team by generating a discussion initiator.
In this scenario, the discussion initiator takes the form of a Meeting Focus Tool that lets meeting participants assess the implications of including specific features in a product from their role perspective and voice their preferences on key features. CoExplorer utilizes GenAI to generate the Meeting Focus Tool--please see Appendix \ref{sec:appendix-B-focus} for the system prompts we used for GPT-4 to generate the tool.
The Meeting Focus Tool is displayed in \autoref{fig:init}B as CoExplorer generates it.
Once all preferences are communicated, as seen in \autoref{fig:init}B, CoExplorer synthesizes these varied responses to refine both the meeting's objectives and phases.
CoExplorer presents the revised information as shown in \autoref{fig:init}C (see~\autoref{sec:appendix-A},~\autoref{fig:phasesSpecific}, right), which narrows the team's focal points for discussion to areas with the most divergent views. Areas of divergence are chosen for discussion because live meetings are most suited to dynamic discussions, including productive conflict, while areas of agreement can be handled asynchronously or set aside for later~\cite{streibel_managersguidetoeffectivemeetings_2003}.
In this scenario, the GenAI system proposed that the tool take the form of a feature ranking aid (see~\autoref{sec:appendix-B},~\autoref{fig:calc} for the initial raw aid generated, and ~\autoref{fig:phasesSpecific} (B) and (C) for the final version). GenAI's versatility could enable the tool to alternately manifest as a chatbot or questionnaire. 

During our experiments, we observed that GenAI is indeed proficient at devising a tool for this need. To enable a later evaluation, we opted to employ a pre-generated version of the tool rather than generating it on-the-fly. This decision was made with the intention of ensuring a consistent experience for all users. Upon utilizing the pre-generated tool, we noted certain aspects of the user interface, such as the color scheme used for highlighting selected buttons, could be misinterpreted. Consequently, we implemented manual adjustments to the pre-generated version to refine the prototype for better clarity and functionality.

\subsubsection{Dynamic Window Management}

Upon examination of the updated phases, the team commences the meeting. CoExplorer curates the necessary documents and applications for each phase, generating an ideal layout for the display. Initially, participants engage in a social introduction phase, where their video feeds are maximized (\autoref{fig:windowingSystem}A). As the meeting progresses from casual conversation to its formal agenda, CoExplorer detects the shift to a new phase, specifically the project introduction. The project manager outlines the problem being addressed.

\textbf{Implementing Design Strategy 1:} To process spoken dialogue, we segmented speech into discrete utterances based on pauses.
For each utterance, a transcript was produced using the Microsoft Azure Speech API.
These transcripts were then provided to GenAI, which determined the pre-identified phase of the meeting the utterance pertained to.
If GenAI's prediction indicated a new phase that differed from CoExplorer's current phase, the user interface was updated to reflect the new phase, shifting to the window layout associated with it. This layout switch was based on a pregenerated list of window configurations established at the start of the meeting.

In organizing the layout of multiple windows, we employed a tiling approach. This decision was influenced by our experimentation with GenAI's ability to generate window sizes and positions (see \autoref{sec:appendix-B-layout} for the system prompts we used to generate layouts). While GPT-4 was adept at creating well-fitted window layouts, GPT-3.5 — the version available to us during prototyping — fell short in this capacity. Consequently, a freeform window layout was deemed unsuitable.
Thus, we established that a tiling window layout optimally utilized the available screen space within CoExplorer. %

\textbf{Implementing Design Strategy 3:} At this juncture, CoExplorer asks all participants on whether they want to halt the transition (\autoref{fig:windowingSystem}C). If no objections occur, CoExplorer adapts the screen layout to fit this more work-focused phase: video sizes are reduced, with a PowerPoint presentation occupying the left side of the display, and a collaborative notepad on the right (\autoref{fig:windowingSystem}B).

In the scenario through which participants were guided, a hardware engineer begins discussing the importance of Bluetooth 5.0 for the headphones. This is taken up by a software engineer who points out the challenges in supporting this feature through software.
Once more, CoExplorer senses a phase shift (requesting a confirmation on whether to proceed with the change; \autoref{fig:windowingSystem}C).
With no opposition, CoExplorer transitions to a ``Discussing Bluetooth 5.0'' phase (\autoref{fig:windowingSystem}D).
The PowerPoint is replaced with an Excel sheet on the left side, a calculator for the Meeting Focus Tool atop the right side, and the notepad downsized to the bottom right, guiding participants back to the highlighted contentious topic.
A full meeting would proceed this way to a decision and discussion of next steps. Due to time limitations, we did not complete the full meeting scenario in CoExplorer. Current GenAI systems such as Microsoft Copilot are able to detect and outline action items from a transcript, and thus a future CoExplorer-like system could use this in conjunction with adapative windowing to place action items under the video of the relevant person.

\section{User Study Methods}
\label{sec:methods}

We organized 26 study sessions as controlled simulations designed to accurately capture experiences at the appropriate moments. One participant joined a researcher in each of these sessions.
The researcher introduced the scenario and interacted with the participant via a semi-structured interview and think-aloud session. 
This approach enabled us to gather more detailed insights related to our research questions than we could have obtained through conventional dyadic or triadic studies.

\subsection{Participants}
We aimed to capture a wide range of experiences with CoExplorer, and thus 
recruited participants with a variety of experiences, levels of expertise, and roles.
Participants were 26 employees drawn from a large global technology company. Employees were recruited through a combination of convenience sampling, snowballing, and batch emails, ensuring a diverse profile by age, region, and gender (Age: 27\% 18-29, 54\% 30-44, 19\% 45-59; Region: 69\% UK (including EU), 27\% US, 4\% Canada;18 males and 8 females). Of 26 participants, 23\% of the participants had the job role title ``Principal'', and 42\% had ``Senior''. Participants included diverse roles such as Project Manager, Cloud Advocate, Software Engineer, IT Service Manager, Hardware Engineer, and Design Researcher. After grouping roles into five broad categories, we had 12\% designers, 4\% hardware engineers, 31\% project managers, 23\% researchers, and 31\% software engineers. %

\subsection{System}
Participants used their personal laptops with a Windows executable of CoExplorer. 
The study proceeded with a Node.js server and a CoExplorer Windows executable which synchronized the state, regulated video communication and had a browser-like windowing system. %
The web page of ChatGPT was manually overlaid with UIKit\footnote{https://getuikit.com}
.
The adaptations to the client application (including connectivity with the client application) were manually added for this user study version of CoExplorer to prevent context size overflow (because we forced ChatGPT to generate 30+ feature-price pairs.).
The interactive phases were manually initiated by the researcher to prevent misclassification of phases based on user responses to interview questions during the study.
To ensure a consistent starting point for all participants, the initial list of phases and a Meeting Focus Tool were pre-generated using ChatGPT. These decisions were made to standardize the CoExplorer experience for the user study.%

\subsection{Study Process}
Each study session lasted approximately one hour and was conducted via Microsoft Teams. Our scenario (discussed above) was chosen from among several that we piloted (e.g. predicting how public policies might affect property prices, and devising designs for a new keyboard) as the most widely relatable to our participant pool.  The following steps illustrates the user study:
\begin{enumerate}
    \item Participants chose a role in the scenario closest to their actual work experience (options included Program Manager, Software Engineer, Hardware Engineer, Designer, and Researcher). This role categorization assisted in prompting realistic experiences.
    \item Participants were introduced to the meeting scenario and CoExplorer. To simulate the meeting lifecycle, participants pasted a calendar invitation into CoExplorer (§\ref{sec:coexplorer}). 
    \item Participants experienced the initial phase list presentation and responded on the focus tool to express their thoughts about the meeting. To demonstrate CoExplorer's ability to highlight differences in opinion among participants, we assigned contrasting opinions to the other simulated attendee roles. Based on differences, phase list refinements happened and the system displayed the refined phase list to participants. 
    \item Participants experienced the simulated meeting. This meeting included social introduction phase within a meeting setting, and a minimum of two different phase transitions. They interacted with the system and discussed the value, challenges, and necessary human involvements regarding each facet of CoExplorer.  To stimulate authentic feedback, the participants were kept unaware of certain implementation details, such as the source of files and applications.
\end{enumerate}

Participants' insights were sought after the study session. Each participant received a %
gift card as thanks.

\subsection{Analysis}
The data collected from the semi-structured interviews and think-aloud sessions were transcribed using Microsoft Teams and later verified by the researchers. This data underwent a twofold analysis process entailing closed coding for investigating attitudes and perceptions towards specific concepts, such as the system choosing files to share, followed by iterative open coding to uncover rationales behind attitudes or perceptions concerning other parts of the prototype.

\section{Findings}

\label{sec:findings}
\subsection{Participants' current meeting experiences}
Participants in our sample frequently participated in remote meetings; 22 daily, nine exceeding ten hours weekly. 
Except for P12 and P18, all engaged in multi-disciplinary meetings.

\paragraph{Meeting Structure}
Over half of our participants (17 of 26) mentioned using predefined agendas for their meetings.
P11 stated this was important to their meetings: \textit{``We always have an agenda and a flow…super important.''}. 
However, 11 participants reported also experiencing non-linear meetings, which was predominantly common among senior roles, like people with `Principal' (4 of 6) in their title, as well as particular roles like project managers (5 of 8) and designers (2 of 3). P22 details this non-linearity: \textit{``very few meetings follow the linear path that one had hoped, and most often you would start with setting up the context and people would have more interest or questions in certain parts of the overall context than others.''}

Eight participants reported meetings to often exceed set time, while nine indicated punctuality, mainly attributed to back-to-back meeting schedules. 
Opinions on meeting durations were divided, with 12 reporting satisfaction and nine reporting dissatisfaction. 
Key challenges included biased coverage of topics (P6, P13, P17, P20, P23), insufficient pre-meeting preparation (P15, P19, P20), and running over time (P12, P13, P19, P22). 
Proposed solutions included pre-meeting voting (P8, P13) and pre-reads (P4, P12, P15, P26).

\paragraph{Multi-window usage}
Most participants (22 of 26)  used multi-window setups for work, consciously arranging the windows, as described by P10: \textit{``My communication tools…on one screen, my main tools…on the main screen''}. 
Despite the popularity of multi-windows in work, 18 participants considered it unsuitable for meetings due to difficulties in screen sharing and inconsistent resolutions, emphasizing setup inefficiency (P7, P9, P10) and issues with varying display sizes (P8, P9, P17, P20). 
These demonstrate a mismatch between user desire and technological support resulting in infrequent use of multi-window setup during meetings.

\subsection{RQ1: Using team input to drive meeting purpose}
CoExplorer provided ways for meeting organizers to capture initial opinions from attendees. 
This has potential to democratize the process of affecting the meeting flow, ensuring that voices are heard while maintaining efficiency.

Almost all participants (24 of 26) found capturing meeting attendees' preferences in advance to be helpful, appreciating the ability to align and prioritize discussion topics. 
P10 mentioned: \textit{``a lot of the discussion often talks about the differences I guess. So it would be good to see where everybody's aligned and where everyone's not aligned.''}
Over half of our participants (17 of 26) wanted this tool well in advance of the meeting to treat it as a preparatory exercise. 
Additionally, six participants suggested adding more ways to express opinions, such as text boxes, to make the focus tool more flexible.

Just over half of participants (14 of 26) appreciated the structured and automatic democratization of meeting flow, as it helped ensure that discussions were focused and avoided repeating topics which everyone agreed upon. 
Participants recognized the potential to save time and delve into more in-depth conversations. 
P18 mentioned, \textit{``I think that's [the concept of the meeting focus tool] a really good idea. (...) We apparently all agree that this [item on the list] is something that we want. And so we can discuss the other two items more in depth.''} 
Eight participants suggested no need for a human mediator for the meeting flow refinement process. 
As P2 mentioned: \textit{`` I think the system can automatically monitor it. Why use human?''}
Additionally, eight participants suggested that the meeting phase list should be refined during the meeting as well. 
They desired real-time phase changes based on meeting outcomes and time management considerations, as P23 mentioned: \textit{``So I would really appreciate if it could change it [the phase list during the meeting] and they could also tell me like you have this much time left you can talk about this.''}

Six participants further emphasized the value of capturing initial thoughts and facilitating collaboration, when asked for the overall advantages of CoExplorer.  
P19 mentioned: \textit{``The biggest win was the creation and synthesis of data into a way that could be consumed in a unique space for collaboration. (...) I loved how the discussion could be reflected in the agenda and the shifts around that and the ability to manage that and even to do collaboration on that. I thought that those were sort of really, really interesting parts and I think it's very interesting to think about how one can manage.''}

\subsubsection{Challenge: Embraced role of organizer}
Contrary to our design, just over half of our participants (14 of 26) wanted the meeting phase refinement results to go through the meeting organizer. 
These participants believed that the organizer would have a better understanding of the meeting dynamics and should have the final say on %
suggestions. 
P8 mentioned, \textit{``The person who's chairing the meeting should be able to accept, reject, or modify any suggestions because they might know better or they might like the suggestion or not.''} 
Participants suggested additional roles for organizers, such as confirming with participants or scheduling follow-up meetings, to enhance meeting management.

\subsubsection{Challenge: Disruptive refinements}
Although many participants appreciated the phase list refinement feature, six opposed pre-meeting refinements, and eight were against during-meeting refinements%
. 
Participants highlighted concerns about the potential disruption caused by refined phase lists. 
Almost half of our participants (11 of 26) worried that the phase list could become too focused, disregarding important topics or overwriting the organizer's agenda. 
P21 expressed their concerns, stating, \textit{``Before the meeting happens, if this (refined phase list) is showing up as if this is replacing what I have manually set as the agenda, then without seeing that, that's strange. Right? Why would it just get overwritten? My stuff gets overwritten.''}

During meetings, P17 was apprehensive about system-driven changes leading to dominance or biased conversation control: \textit{``If someone has a dominant personality, they can take over the meeting. That might be made even worse if someone was speaking a lot and then the phases changed based on that.''} 
P4 and P9 raised concerns about refining phases during meetings without the right stakeholders present. 
P9 mentioned, \textit{``I wouldn't want the phases to change in the meeting because again, the meeting is called with specific participants. If you had a phase change during the meeting, would the right stakeholders be there? I would rather it take the outcomes or the output of the meeting and then suggest further meetings.''}

\subsection{RQ2: Surfacing implicit needs and resources}
CoExplorer allows participants to structure their meetings without adding additional tasks to their existing meeting routines through automated phase suggestion and refinements. 

Seven participants appreciated the easy structuring capability provided by CoExplorer. 
P10 mentioned, \textit{``So from just a line of text to get a basic structure is very quick, and having someone then able to tweak that is really, really useful.''}, and P11 \textit{`` I really, really like the first feature that you show me how using an AI model based on a meeting description can generate the agenda items. That's super cool. The second one, the previous input meeting where participants select some topics [in the meeting focus tool] and based on that, adjust the agenda. Those two were mind blowing. Well, those two for me was, oh my God, this is a great way to pick up something big.''} 
Participants' comments emphasized that CoExplorer showed potential to improve the following effectiveness issues:
\begin{itemize}
    \item \textit{Productivity of meetings} (P10, P16, P22, P25): P22 said \textit{``Too many meetings just don't achieve anything for anyone just because the meeting organizer did not prepare the agenda did not even know what the agenda should be. The flow should be what things to take care of, how to modulate the discussion, how to move from a phase to phase. Hard stuff. So support is definitely needed.''}
    \item \textit{Clarity of meeting goals} (P8): \textit{``Much better clarity on whether the meeting goals are being met''}
    \item \textit{Efficiency of meetings} (P10, P23): P23 said \textit{``I think I can speed up things because whenever I do meetings, I'm like, OK, give me a second. Let me find that document. Ohh yeah, here's the relevant document or something like that.''}
\end{itemize}

Four participants %
also valued CoExplorer's ability to assist meeting organizers in driving the meeting. 
Such a system is particularly helpful for meeting organizers given the tendency for people to impose most responsibility on the organizers to be the primary drivers of meetings. 
People saw it as a useful tool for meeting orchestration, especially for organizers who may struggle with managing complex cross-functional meetings, as mentioned by P22: \textit{``I think what you're trying to do here makes a lot of sense. And in my experience, I would not assume that more than 20\% of people who hosted meetings like these feel comfortable and confident of orchestrating these tricky conversations.''}

\subsubsection{Challenge: Capturing nuances}
Participants expressed concerns about the system's ability to detect nuances in the files and conversations they create. 
Eight participants mentioned having multiple versions %
of files while working and struggled to see how the system could accurately infer which version to use (e.g., private vs. public versions, variations on ideas, and iterative refinements). 
They relied on visual representation or location cues to differentiate files. 
P5 stated, \textit{``I don't really number the versions really well. I'm trying to do like V1, V2. But then the way I remember which files to share and which files not to share is by remembering their location. So, I need that visual representation of where each file is.''}

Participants also stressed the challenge of system detection and reaction to subtle nuances in the conversations during meetings. 
Some participants (P3, P4, P6, P7, and P22) believed that the system should consider the emotions and thoughts of individual attendees when refining phases. 
However, participants acknowledged the difficulty of quantifying and addressing these nuanced aspects. 
P22 questioned whether it was feasible or desirable for an AI system to index on emotions and undertones, stating, \textit{``Alongside the rational content of the discussion there are some undertones inevitably here and as an orchestrator we try and respond to that. I don't know if that can be or even should be replicated in an AI system here. Maybe we are better off just getting rid of all those emotions and undertones. Maybe, maybe not.''}

\subsubsection{Challenge: Excessive structuring of the meeting}

While many participants appreciated the provision of structure by CoExplorer, some (P1, P4, P9, P12, P20, and P22) voiced concerns about excessive structuring. 
P9 expressed concerns about the impact of %
imposed time limits: %
\textit{``It rather limits the discussion or can limit the discussion. If we say, hey, discussing target users but no more than 15 minutes, this is never enough to discuss target users. What if we end up spending 30 minutes on it? Do we then not talk about hardware features?''} 
P22 suggested reducing phase level detail: \textit{``Rather than going into six different phases and names and descriptions of that, that felt like it is putting too tight a structure already. Instead, if it stayed at the level of just doing three different phases broadly and one liner descriptions, that's a good enough starting point I felt.''} 
P12 noted that overly structured meetings would not align with their company's norms. 
P12 said, \textit{``We probably wouldn't pay too much attention. ... There may be organizations where meeting culture is much more regimented and they do look to things like this to structure their agenda.''} 
Overall, these thoughts from participants indicate the value and challenge of non-linear meetings. 
Conversations often go off on tangents and some people value this either implicitly (e.g., P4: \textit{``I think it could create very linear meetings.''}, implying that linear meetings might be undesirable) or explicitly (e.g., P1: \textit{``If that meeting is structured, then people might feel pressured to only stick to the point''}). 
This suggests the importance of balancing meeting structure and flexibility in meetings.

\subsection{RQ3: Merits and challenges of implementing Human-on-the-Loop (HOTL) interactions}
We used the HOTL interaction paradigm to manage changes in phases, and the windowing system to minimize distractions and information overload during meetings. 
Participants had mixed responses to the various aspects of HOTL design, and we identified some common challenges.

Ten participants expressed excitement about the system's capability of curating shared applications and files. %
P4 appreciated the system's ability to improve document discoverability, allowing participants to easily access relevant resources during meetings: \textit{``One of the most difficult things in meetings is, do we have a document around this... So having that as resources that people can get to easily is really vital in a meeting.''} 
P4, P17, and P24 mentioned that system could auto-populate the windows without verifying the window `sharing' decision with any attendees. 
Furthermore, almost half of our participants (12 of 26) found that the system showing content in the shared task space was not intrusive.

Participants also provided input on the search scope for curation. 
A slim majority (15 of 26) favored curating work-related files, such as those within a shared work cloud, as they assumed team members would have common access permissions. 
In contrast, indexing files from the internet (4 of 26), local systems (6 of 26), or anywhere available (2 of 26) were less preferred options. 
Additionally, over half of our participants (17 of 26) mentioned trusting the curated contents. 
However, many participants linked their trust to conditions, such as their continued experience with the system (P2, P21, P23, P25) or the ability to see previews of the chosen files (P3, P4, P10, P12, P26).

The ability of the system to automatically add windows was positively received by almost half of our participants (12 of 26; 7 were neutral). 
Participants found it particularly useful for quickly retrieving and sharing information during meetings, thereby making meetings more efficient. 
P16 mentioned: \textit{``I think in terms of getting information out quickly, rather than having to look for it all, it would be useful. I feel like a lot of meetings they're a big time sink. It's just people switching in between apps.''} 
Six participants also appreciated the concept of the system resizing and relocating windows, finding it beneficial for organizing and managing their workspace during meetings.

However, participants were less inclined to have the system remove windows. 
Eight expressed that they did not want the system to close windows, but most (20 of 26) mentioned that they expected the system to allow for window recall in a format similar to the last opened window interface in web browsers or a hamburger menu.

\subsubsection{Challenge: Agency}
Using the HOTL %
paradigm, we only allowed users to \textit{abort} actions performed by the system. 
This raised concerns among participants, who desired a more traditional human-\textit{in}-the-loop approach to retain agency%
.

Most (21 of 26) participants mentioned that the system should consult humans when showing any file or application on the shared task space. 
Additionally, nine participants believed that similar checks should be performed for windowing system operations, and over half (16 of 26) said that checks should be performed for phase changes.

Participants recognized the potential distractions that such prompts could cause during meetings. 
They emphasized the need to reduce the frequency of prompts and suggested offloading these prompts as a pre-meeting exercise. 
P6 highlighted the necessity of reducing the frequency of prompts: \textit{``You don't want to present and then you have something popping up quite regularly for some reason. That could be distracting.''}, as well as P13: \textit{``I wouldn't want to be in a meeting and then be notified 20 times that we're now moving to a next slide.''} 
Six participants suggested the creation of a ``Meeting choreography'', where the meeting organizer reviews and decides on the files, applications, and phases that should be prepared before joining the meeting. 
Similarly, nine participants proposed pre-defining the scope of the search for curation, such as putting files into a certain folder or tagging files in advance of the meeting.

\subsubsection{Challenge: Algorithmic aversion/mistrust}
Participants expressed concerns about the ability of the system to generate content accurately.
Nearly every participant expressed significant skepticism and mistrust regarding the actual content generated by CoExplorer without seeing the system err (except P20 who mentioned that some of the suggestions given by CoExplorer were not suitable). This suggests of a pre-existing aversion towards, or mistrust of, automated systems.  
As P9 summarized, \textit{``My experience with meeting technology is when people walk in and they click go and it works, they are super happy. The minute something doesn't work, they are super unhappy, and there's no middle ground.''} 
Participants emphasized the importance of the system understanding the nuances and context correctly, without displaying irrelevant screens or inappropriately transitioning the phases, based on their experience with meeting technology. 
In particular, seven participants criticized the system's inability based on their existing experiences, sometimes even before using the system, rather than evaluating the system's performance during the study. 
P3 based their view of the system on their experience with ChatGPT: \textit{``So a lot of these multidisciplinary meetings are quite sensitive. They're quite delicate balance of understanding each other's perspective so people don't get hurt.(...) Expecting ChatGPT to come up with all these subtle nuances and understand these nuances, it's slightly unrealistic because that's never going to happen.''} 
P5 also expressed concern regarding an algorithm-generated decision: \textit{``I feel like if this is auto generated, I might not even know if this is something that we're gonna talk about in this particular meeting for now.''}
These concerns highlighted the need for the system's content generation to be dependable and aligned with participants' expectations.

\section{Discussion}

\label{sec:discussion}
Our research provides empirical early evidence for design concepts (\autoref{sec:designconcepts}) that can enhance the productivity of meetings. There is value in GenAI's ability to extract and reformulate goal-direction from existing information to support intentionality in meeting planning and execution. This includes helping organizers highlight meeting goals and adjust plans to address inter-attendee issues before the meeting, and surfacing the potential implicit phase structure of the meeting in advance, while adapting it to in-the-moment activities. Our design concepts were perceived to be more useful when larger quantities of resources and participants needed management, and when the meetings were intended to be more structured than non-linear. Participants gave us insight into circumstances where GenAI might not be beneficial, and their overall concerns about the place of and trust in GenAI for augmenting meeting intentionality.%

\subsection{Trust in AI systems: balancing system autonomy with user trust}
Participants appreciated our intention to equip meeting attendees with an intelligent system.
However, despite the potential benefits the study shows, there is a general reluctance among meeting attendees to trust AI systems which do more than merely give suggestions. 
This resonates with existing literature on algorithmic aversion within decision-making fields \cite{dietvorst_algorithm_2014}. 
Participants wanted decisions made by the system to be validated by the meeting organizer. 
Even though CoExplorer did not make mistakes in predictions for most participants, we still noted a marked mistrust toward CoExplorer.
This issue surfaced in the initial stages of the user study when interaction with the system was limited, and lasted throughout.

Designers need to address this mistrust to integrate more advanced intelligent systems into meetings. 
The majority of participants and existing literature suggest a human-in-the-loop approach \cite{xia_crosstalk_2023, pu_semanticon_2022}, highlighting the need for human intervention and oversight. 
However, frequent interventions can overload meeting attendees with information, reducing their capacity to process it. %
Further, constant system prompts could be disruptive and distracting to the meeting process, as recognized by our findings.
Even sharing a single file could involve reviewing multiple pieces of information---location, permissions, format---which makes it pivotal to balance the trade-off between information provision and disruption.
Future investigations could focus on identifying the threshold of disruption for informational prompts---how much information is it safe for these prompts to have, and how frequently can they be shown to be effective? 

\begin{figure}[h]
    \centering
    \includegraphics[width=0.45\textwidth]{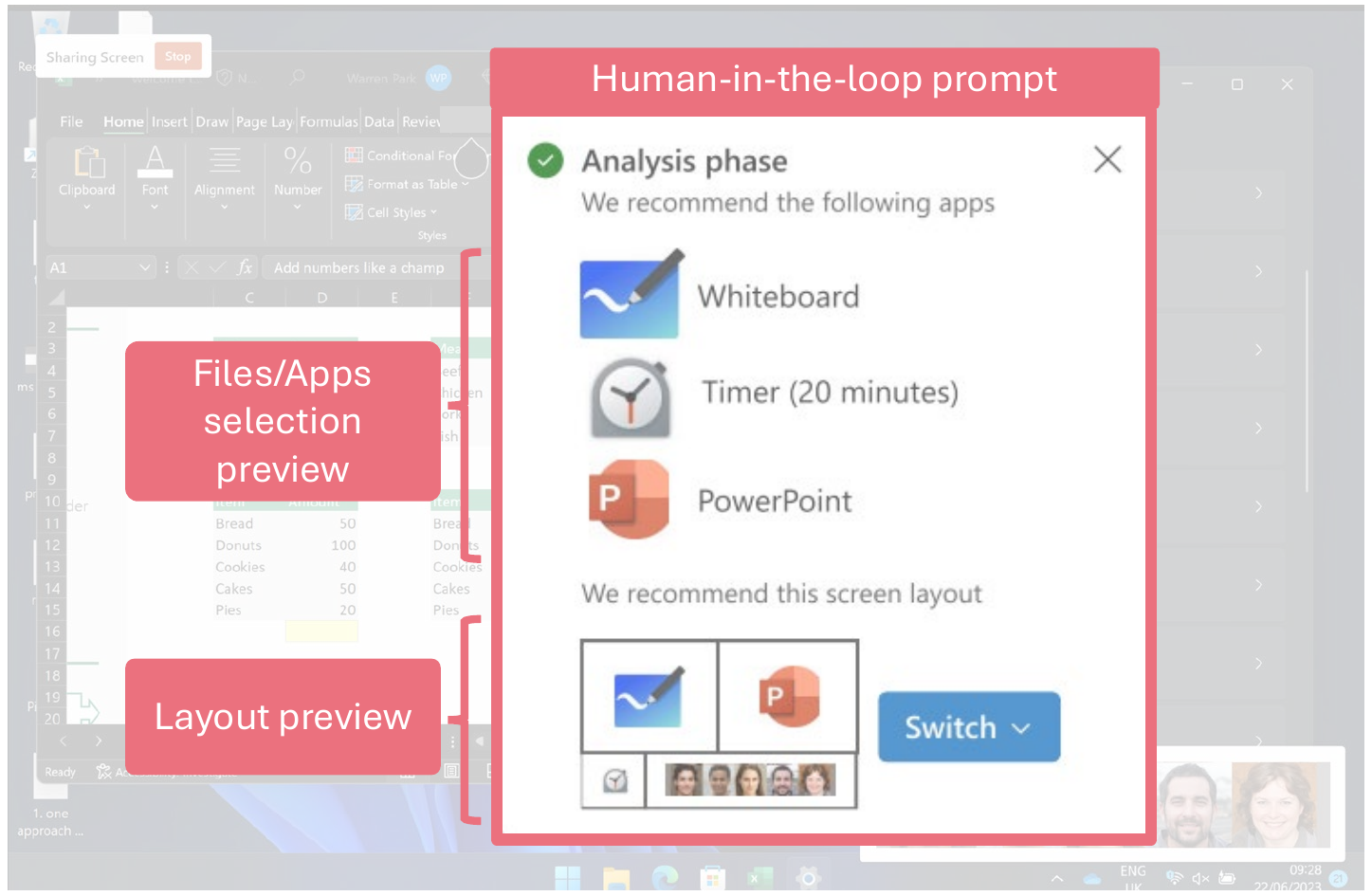}
    \caption{An example human-in-the-loop design for verifying app selections}
    \Description{Shows an example screen. There is a human-in-the-loop prompt in the middle of the screen which shows the recommended layout for a phase. This recommendation consists of a files/apps selection preview and a layout preview.}
    \label{fig:appSelector}
\end{figure}

\vspace{2mm} %
\noindent\fbox{%
   \parbox{0.45\textwidth}{
\textbf{Design implication 1:} A balance between keeping meeting attendee effort low during the meeting, while offering more agency and establishing trust, could be reached by off-loading some of the human-in-the-loop tasks to periods of low engagement---prior to the meeting, during breaks, or when certain attendees are less engaged. Example human-in-the-loop prompts are shown in \autoref{fig:appSelector}. 
}
}

\subsection{Structure in meetings: linearity, stages, and approach}

\subsubsection{Linearity of meetings}
Our findings suggest that both linear and non-linear meetings are common for our study participants, leading to mixed responses about surfacing the implicit phase structure in meetings. 
Some participants valued the structure that CoExplorer imposed on meetings by surfacing meeting goals and phases, helping to clarify attendee perceptions in advance of meetings, and supporting organizers in driving the meeting. This was perceived to improve meeting clarity and therefore productivity. However, some participants expressed concerns about the excessive structure imposed by CoExplorer, worrying that it may limit the natural spontaneity and flexibility of conversations. 
Considering these individual perceptions and preferences, it may be crucial for meeting support systems to capture the general consensus on the desired level of structure and adapt the workflow accordingly, rather than adopting a one-size-fits-all approach. 
Reconciling the variations in the group consensus and the individual preferences is a direction that needs further research.
Systems could also help people decide on how much structure may be needed depending on their goals.
It also points out the value in developing ways to adaptively adjust the level of structure during meetings. 
For instance, when people are moving through phases smoothly, the suggestion of structure could be more restrained, whereas if the meeting is being side-tracked, structure could be suggested in a stronger way by tracking the progress of the meeting through an intelligent system.
\vspace{2mm} %
\noindent\fbox{%
   \parbox{0.45\textwidth}{
\textbf{Design implication 2:}
The design of intelligent meeting systems should consider ways to address the tension between structured, linear meetings enabling intentionality and clarity, and less structured, non-linear meetings, enabling flexibility. This includes capturing attendees' preferences about structure as well as helping decide how much structure may be beneficial for their goals. Variations in need for structure during a given meeting should also be considered to adapt the interface.
}}

\subsubsection{Meeting stages and broader context}

Our study underscores the strengths of intelligent systems in creating a trajectory of intentionality across the meeting lifecycle. %
This can help workers focus on making meetings matter by augmenting effort rather than hoping that unfocused effort during the scheduled meeting time will have good results.

An interesting avenue for expansion might be giving the systems capability to track progress beyond the context of individual meetings, and tapping into the larger context of team projects. For instance, the meeting focus tool or phase generation could adapt to the stage the project is in (e.g., initial conceptual design, prototyping, or pre-launch). 
This might have several benefits.
For instance, a system that tracks an entire project could prevent the team from wasting time rehashing decisions that were already made. We could bolster user interface support with features such as `graying out' previously discussed topics to signal they should not be revisited. Further research may illuminate how meetings can best use GenAI to ensure they are run more efficiently and coherently.

\vspace{2mm} %
\noindent\fbox{%
   \parbox{0.45\textwidth}{
\textbf{Design implication 3:}
The meeting life cycle extends much further in both directions (before and after meetings), and designing interactions for these stages might make attendees more intentional throughout the meeting life cycle. 
For instance, the meeting focus tool might be most beneficial when shown to attendees well in advance of the meeting. 
We should also consider designing experiences that adapt beyond just the scheduled meeting, such as the context of the larger project within which the meeting is taking place.
}}

\subsection{Task load and agency: striking a balance between organizers and attendees}
Our design intention for CoExplorer was to reduce the effort and task load for meeting attendees and organizers by using automation. 
However, participants' feedback indicates the need for some human intervention, with most pointing to the meeting organizer or chair as the appropriate person to validate the decisions made by the AI system.
For example, participants suggested more human-in-the-loop design elements or more previews for organizers, which provides information on differing options and opinions among attendees. 
The implication is that the organizer, as an obvious target, is responsible for resolving these differences before the meeting begins. 
Implementing such a system, then, reduces the task load and effort for meeting \textit{attendees} at the expense of higher task load for the meeting \textit{organizer}. 
This might be mitigated by ensuring that the added tasks for the organizer happen during low engagement times, as discussed in Design Implication 1 above.

These implied tasks are a consequence of participants' perceived necessity for human moderation of decisions made by GenAI systems. 
Indeed, some participants saw the current design as limiting their agency during meetings. 
CoExplorer's phase suggestions were perceived as a rigid task list that devalued their contributions, and would prefer such a system to be a ``suggester'' rather than a ``dictator''. 
This suggests a tension between the simultaneous desire for low task load (e.g., fewer verification prompts) and high agency (e.g., more human moderation). 
This tension merits further investigation.

\vspace{2mm} %
\noindent\fbox{%
   \parbox{0.45\textwidth}{
\textbf{Design implication 4:}
There's a delicate balancing act between task-load and agency that an intelligent meeting system must manage. Systems should act as an assistant—reducing task-load and enhancing agency—not a decision-maker.
}}

\subsection{Balancing efficiency and sociality}
Per RQ1 (using team input to drive meeting purpose), many of our participants appreciated that the meeting focus tool could help steer meeting activity to the topics that needed most discussion, in this case based on differences of opinion that needed to be resolved. This is clearly a step in the right direction towards effective meetings, but there are other factors at play. First, many meetings have a social component or may be entirely about sociality, such as establishing or maintaining collegiality~\cite{bergmann_meeting__vcfatigue_2022}. To the extent that a meeting is the right venue for sociality, privileging only task efficiency with no time or highly constrained time for small talk at the beginning and end of meetings~\cite{yoerger_chitchatpremeeting_2015} may be counter-productive. It might not only reduce team engagement but also reduce serendipitously raising important issues. Further, if all meetings focus only on points of difference, this might lead to problems with team morale, making it seem as if the team disagrees more than is actually the case, and/or leading to a dread of meetings because they always involve the difficult work of resolving difficulties. There may be value in teams first considering their points of agreement to establish common ground and trust for dealing with disagreement.

As such, even if resolving disagreements is the most important value that a meeting has, either the meeting focus tool or some other aspect of the meeting focus and/or phase creation of the GenAI needs to be guided with meta-prompts to balance sociality and agreement with tasks and disagreement. Gonzales-Diaz et al.~\cite{gonzalez_diaz_makingspaceforsocialtime_2022} propose a version of of this balance in their Transitions concept. The propose an interface that literally changes before, during, and after a meeting. Pre- and post- meeting interfaces are simple blank spaces that afford free-form spatialized talk, while the during-meeting phase changes to look more like a traditional meeting space, in their case an auditorium, which is also a literal visual indication of space for `real work' and a literal focusing mechanism of an audience on a presenter. They note that their design of the visual look and user experience of ``making space for social time'' needs significant work, and this might be quite different for different teams. As such, this would be another area in which GenAI's ability to make use of user preferences might create very different-looking meeting interfaces for different teams while following the guiding principle. The precise balance of effectiveness and sociality will, of course, require future exploration.

\vspace{2mm} %
\noindent\fbox{%
   \parbox{0.45\textwidth}{
\textbf{Design implication 5:}
The design of intelligent meeting systems should consider ways to balance effectiveness with sociality. This must preserve the social and ritual value of meetings when it is needed. It must also ensure that the drive for effectiveness does not create a cascade of social problems that undermine the very effectiveness that is aimed for, such as balancing agreement and disagreement so that attendees can trust that conflict will be productive.
}}

\subsection{Limitations}
We acknowledge that our study has some limitations. Our participants were recruited from a single company. 
This choice was made to maintain a consistent meeting culture across our participants while comparing differing perceptions towards the same prototype. 
Future research could broaden the scope to include multiple varied companies and sectors with differing meeting practices.

Our participant pool was not gender-balanced (female participants were only 31\% of the total, none were non-binary). 
The 31\% female proportion is unfortunately similar to the average in the technology sector in 2023 \cite{techopediaWomenTech}, but future research should seek greater gender balance.

Since confidentiality issues prevented participants from bringing their actual work files into the study, our guided walk-through used fictitious documents. Future GenAI design research will, of course, benefit from incorporating real-world data, documents, meetings, etc., which would allow participants to evaluate the benefits or drawbacks with more relevance to their work.

We also acknowledge that our study method did not validate the generated phases and support tools, nor was it a comparative evaluation of GenAI creating an interface versus an interface created by a person, or GenAI running meetings versus meetings run by people.
These limitations are a natural consequence of choosing the guided walk-through method. 
In common with the Participatory Prompting method~\cite{sarkar_participatoryprompting_2023}, rather than speculating on fictional GenAI capabilities, our technology probe walk-through allowed users to engage with a working example of a system that was not yet possible to build to a fidelity necessary for plausible evaluation.
For example, %
as of the time of writing, GenAI is not proficient at abstract reasoning (e.g., see~\cite{lewis2024using, mitchell2023comparing}), 
which limited the kinds of meeting phases that our implementation of CoExplorer could detect.
Given the early state of GenAI in this usage area, our approach allowed us to capture rich insights into users' reactions to GenAI interventions at various stages of the meeting lifecycle, and what challenges and opportunities that might bring.

Finally, this research did not extensively explore a range of design options for meeting interfaces, as this study focused on users' reactions to the meaning and logic of GenAI driving interface changes, not on evaluating or optimizing the look and feel of the system.
We settled on the current design based on a pilot study of the clarity of the logic for users, 
but we could have experimented with more ways to present and organize different phases and tools through co-design sessions or focus group studies.
We leave such investigations to future work.

\section{Conclusion}
\label{sec:conclusion}

We designed CoExplorer, a GenAI video meeting system that aims to promote intentionality while reducing the effort of participating in a meeting.
We used CoExplorer as a technology probe, utilizing specific design concepts that provoked strong responses from our study participants.
We delved into possible user concerns regarding the automation of meeting planning duties and surfacing implicit needs and resources during a meeting. 
Our research further highlighted the difficulties in maintaining user control amidst the automation of decision-making aspects. 
Notably, we discerned a possible algorithmic aversion or mistrust that could potentially hinder the success of future systems. 
Based on these findings, we discussed the challenges in depth and proposed design implications linked to our results.
Additionally, we also provide the prompts we used with an LLM in order to generate the UI and responses in the Appendix.
Our results and design implications suggest next steps for creating GenAI-mediated meeting systems, and the Appendix helps other researchers recreate and build upon our system.

\begin{acks}
We thank Ava Scott for helpful discussions, and the study participants for volunteering their time. We acknowledge the use of ChatGPT by OpenAI on the implementation of the prototype, as well as proofreading the text.
\end{acks}

\bibliographystyle{ACM-Reference-Format}
\bibliography{references,refs-additional}
\newpage
\appendix

\onecolumn
\section{Appendix A: Phase Generation and Refinements (Higher Resolution)}

\label{sec:appendix-A}
\begin{figure}[H]
    \centering
    \resizebox{0.8\linewidth}{!}{
    \begin{tikzpicture}[node distance=0.5cm]

        \foreach \title/\subtitle [count=\i] in {
            {Introduction (5 minutes)}/{Setting the stage for the meeting, outlining the goal and expected outcomes},
            {Discussing Target User and Price Point (15 minutes)}/{Identifying the target user for the new headphones and deciding on a suitable price point},
            {Discussing Software Features (10 minutes)}/{Discussing the software features that the new headphones should have},
            {Discussing Hardware Features (10 minutes)}/{Discussing the hardware features that the new headphones should have},
            {Discussing Design Features (10 minutes)}/{Discussing the design features that the new headphones should have},
            {Conclusion and Next Steps (10 minutes)}/{Summarizing the decisions made during the meeting and outlining the next steps}
        } {
            \node[draw, text width=8cm, minimum height=1cm, align=center, font=\sffamily] (bar\i) at (0, -\i*1.5) {\textbf{\title} \\ \subtitle};
            \ifnum\i>1
                \node[below=1.5cm of bar\the\numexpr\i-1\relax.south] (gap\i) {};
            \fi
        }

        \foreach \title/\subtitle [count=\i] in {
            {Introduction (5 minutes)}/{Setting the stage for the meeting, outlining the updated goal and expected outcomes},
            {Discussing Bluetooth 5.0 (20 minutes)}/{Discussing the inclusion of Bluetooth 5.0in the design of the new headphones},
            {Discussing Auto Pairing (20 minutes)}/{Discussing the inclusion of Auto Pairing in the design of the new headphones},
            {Conclusion and Next Steps (10 minutes)}/{Summarizing the decisions made during the meeting and outlining the next steps}
        } {
            \node[draw, text width=6cm, minimum height=1cm, align=center, right=2cm of bar\i, font=\sffamily] (barR\i) {\textbf{\title} \\ \subtitle};
            \ifnum\i>1
                \node[below=1.5cm of barR\the\numexpr\i-1\relax.south] (gapR\i) {};
            \fi
        }

        \draw[->, thick] ([xshift=4.5cm,yshift=-0.15cm]gap3.center) -- ([xshift=-3.5cm,yshift=-0.15cm]gapR3.center);

    \end{tikzpicture}
    }
    \caption{List of generated initially generated phases (left) and the refined phases (right). After refinement, the phases became more focused and realistic for the given timeframe.}
    \Description{Shows lists of generated phases (all text).}
    \label{fig:phasesSpecific}
\end{figure}
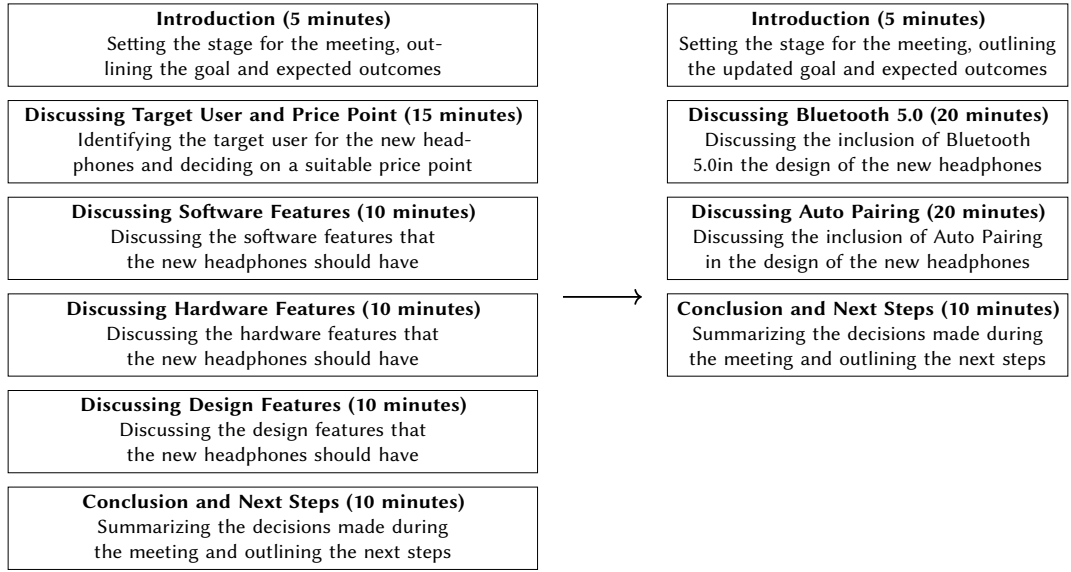

\section{Appendix B: Example prompts}
\label{sec:appendix-B}
We attach versions of the example prompts for GPT 3.5-Turbo that we used for the implementation of the CoExplorer. CoExplorer generated prompts automatically based on the templates. During this, CoExplorer optimized the prompts as per the user inputs, and prompted the language model multiple times to obtain the results in the correct format. The prompts given is one instantiation of output prompts that the CoExplorer was producing.

\subsection{Phase Generation}
\label{sec:appendix-B-phase}

\subsubsection{Example system prompt}

\begin{spverbatim}

[No prose]
[Output only JSON]
Do not write normal text
You are a JSON generator which converts meeting agenda text into a more descriptive agenda description. You always need to have an introduction phase at the beginning.
\end{spverbatim}

\subsubsection{Example user prompt}
\begin{spverbatim}
Please breakdown the following meeting agenda that someone has sent in email into meeting phases that we would need to go through. We have a 60 minute meeting scheduled. Based on the given information, give a goal of the meeting (goal), as well as the explanation on why you chose the goal (exp). And also give the phase definition in a list (pi). Each phase definition should include: (1) Phase title (pt), (2) Phase description (pd) which should include a sub goal of a phase, (3) Behaviors to be encouraged (be), (4) Behaviors to be discouraged (bd), (5) priority (p), (6) amount of time allocation (t), (7) direction (d) (i.e., is it an iterative phase or directional phase). Please only respond in JSON with each element needed as a key within a phase i.e., if we have two phases,
[{"pt":"xxx","pd":"xxx","be":["xxx","yyy"],
"bd":["xxx","yyy"],"p":"high","t":2,"d":"iterative"},
{"pt":"yyy","pd":"zzz","be":["ttt","kk"],
"bd":["lll","mmm"],"p":"low","t":8,"d":"directional"}].
So the overall JSON to export is {"goal":"xxx","pi":
[<phase definitions>],"exp":"xxx"}. Explanation should start by saying this goal is generated... or similar. Please use the full 60 minutes. Here is the meeting invitation:
\end{spverbatim}
(The email invitation would be attached.)

Refinement could be done using the same script, but by attaching the refinement scenario. This script can be generated through code, and does not have to be detailed.

\subsection{Layout Generation}
\label{sec:appendix-B-layout}

\subsubsection{Example system prompt}
\begin{spverbatim}
[No prose]
[Output only JSON]
Do not write normal text
You are a helpful assistant who creates screen layout that has appropriate apps that are most helpful for users to complete the task successfully. Respond only in JSON following the format. Example format:
[{"PhaseTitle":"xxx","timer":n,"programList":
[{"name":"yyy,"description":"zzz"},
{"name:"kk","description":"lll"}]},
{"PhaseTitle":"xxx","timer":n,"programList":
[{"name":"yyy,"description":"zzz"},
{"name:"kk","description":"lll"}]}]. 
Strictly follow this format. n is integer, and 
programList.name should either be a name of a program in a program list given or a URL. programList.description is where you should put a extremely concise reason why you chose that program. e.g., Use this for presenting agenda; Use this for viewing relevant budget data
\end{spverbatim}
\subsubsection{Example user prompt}
\begin{spverbatim}
    I will give you the list of phases in a meeting in JSON format. Each phase in JSON is defined with several keywords. "pt" represents phase title, "pd" represents phase description, "be" represents behaviours to be encouraged, "bd" represents the behaviours to be 
    discouraged, "p" represents priority (high, medium, low"), "t" represents recommended duration of time for the phase, and "d" represents directionality (directional i.e., cannot be returned, and should be preceeded by a certain phase or iterative i.e., can be transitioned to this whenever). You need to generate what kind of programs are needed for helping goals of each phase (defined by the description) to be met the most 
    efficiently. You can generate a list of 1-5 program name/URL, and the sequence of generation will affect where they are being placed, and size. Therefore, you need to be sensible about ordering so that important programs can be shown with the bigger sizes. The rule is as follows: If you have one program on the list, that would be full screen. If you have two programs, it would be one on the left half (first program on the list), and one on the right half. If you have three programs, the right-hand-side panel will split in half, creating two small panels at the top and the bottom. Four programs mean the left-hand-side panel will also be split. Five programs mean two equally sized panel at the top, and three equally sized panel at the bottom. The ordering in the list will be used to place program to panels in a clockwise ordering (top left panel is the first panel). Here is the list of programs available, and if the program that you want is not listed, please generate a URL for the program that you need instead of the program name. Please feel free to give Bing Search URL with the search term filled, and generate at least one URL:

\end{spverbatim}
List of programs could take arbitrary formats.

\subsection{Meeting Focus Tool}
\label{sec:appendix-B-focus}

\subsubsection{Example system prompt for a specialized calculator generation}
\begin{spverbatim}
   You need to generate a HTML page, and only HTML+CSS
   +JavaScript based page as a response which allows me to calculate the total value to have the a list of features for a given scenario. Each feature needs to have an "include" button (green tick) or "exclude" button (red cross), and you need to calculate the total at the end, when the submit is clicked. You should not show any prices including itemised ones before this. You need to generate at least 30 features (features should be unique and descriptive. No Feature 26 or something like that), and incorporate that into the page, those are relevant for the scenario. Assign random prices for each feature. The HTML page needs to incorporate all the features that you generated embedded (i.e., no ... or "many more features here" etc.). No prose. No add more features here etc. You need to list all the features on the HTML. 

   
\end{spverbatim}

\subsubsection{Example user prompt}
\begin{spverbatim}
   Designer, software engineer, hardware engineer, PM, marketing expert, and a researcher are gathering to think about what features that a new headphone product that they release might have. It could be electronics feature such as active noise cancelling (95%
\end{spverbatim}

This user prompt may not occasionally result in explicit features being stated but rather like \textit{``<!--Here will be inserted many more features-->"}. Such cases can be detected, and additional user prompt could be given, for instance:

\begin{spverbatim}
    It does not show 30+ features
\end{spverbatim}

This generates the following working web page:
\begin{figure}[h]
    \centering
    \includegraphics[width=0.3\textwidth]{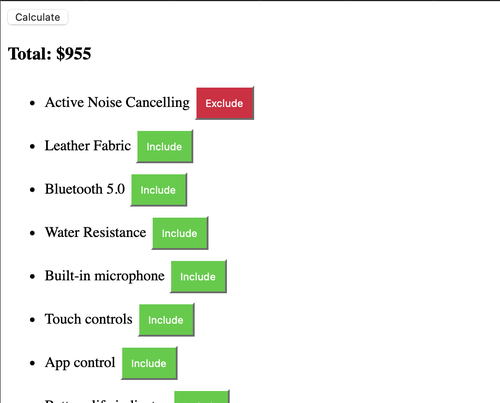}
    \caption{Example raw ranking aid app generated by ChatGPT as a meeting focus tool}
    \Description{Shows an example web app. The top of the app shows the total price, and below it, it shows a list of features and buttons to select whether to include or exclude the feature.}
    \label{fig:calc}
\end{figure}

\end{document}